\documentclass[conference]{IEEEtran}
\IEEEoverridecommandlockouts
\usepackage{caption} 
\usepackage{soul}
\usepackage{lipsum}
\usepackage{indentfirst}
\usepackage{booktabs}
\usepackage{enumerate}
\usepackage{tabu}
\usepackage{tabularx}
\usepackage{environ}         
\usepackage{etoolbox}        
\usepackage[dvipsnames]{xcolor}
\usepackage{eurosym}

\usepackage{cite}

\usepackage{graphicx}
\usepackage{wrapfig}
\usepackage[labelformat=simple]{subcaption}

\usepackage{float}

\usepackage{amsfonts, eqnarray, amsmath, amssymb}
\usepackage{mathptmx}
\usepackage{mathtools}
\usepackage{xfrac}
\usepackage[cmintegrals]{newtxmath}
\usepackage{breqn}
\usepackage{bm}
\usepackage{soul}
\usepackage{environ}

\newlength{\myl}
\expandafter\let\expandafter\origequation\csname equation*\endcsname
\expandafter\let\expandafter\endorigequation\csname endequation*\endcsname
\long\def\[#1\]{\begin{equation*}#1\end{equation*}}
\RenewEnviron{equation*}{
  \settowidth{\myl}{$\displaystyle\BODY$} 
  \origequation
    \ifdim\myl>\linewidth
      \resizebox{\linewidth}{!}{$\displaystyle\BODY$}
    \else
      \BODY 
    \fi
  \endorigequation
}

\makeatletter
\def\@begintheorem#1#2{%
    \parskip 0pt 
    \trivlist
    \item[%
        \hskip 10\p@
        \hskip \labelsep
        {{\bfseries #1\hskip 5\p@\relax#2.}}%
    ]
    \it
}
\def\@opargbegintheorem#1#2#3{%
    \parskip 0pt 
    \trivlist
    \item[%
        \hskip 10\p@
        \hskip \labelsep
        {\bfseries #1\ #2\       
   \setbox\@tempboxa\hbox{(#3)}  
        \ifdim \wd\@tempboxa>\z@ 
            \hskip 5\p@\relax    
            \box\@tempboxa       
        \fi.}%
    ]
    \it
}

\makeatother

\usepackage{algorithm}
\usepackage{algpseudocode}

\makeatletter
\newcommand*{\algrule}[1][\algorithmicindent]{%
  \hspace*{.2em}
  \vrule 
  \hspace*{\dimexpr#1-.2em-.4pt}%
}

\newcommand{\StatePar}[1]{%
  \State\parbox[t]{\dimexpr\linewidth-\ALG@thistlm}{\strut #1\strut}%
}
\renewcommand{\ALG@beginalgorithmic}{\offinterlineskip}

\newcount\ALG@printindent@tempcnta
\def\ALG@printindent{%
  \ifnum \theALG@nested > 0
    \ifx\ALG@text\ALG@x@notext
    \else
      \unskip
      \ALG@printindent@tempcnta=1
      \loop
        \algrule[\csname ALG@ind@\the\ALG@printindent@tempcnta\endcsname]%
        \advance \ALG@printindent@tempcnta 1
        \ifnum \ALG@printindent@tempcnta<\numexpr\theALG@nested+1\relax
      \repeat
        \fi
    \fi
}
\patchcmd{\ALG@doentity}{\noindent\hskip\ALG@tlm}{\ALG@printindent}{}{\errmessage{failed to patch}}
\makeatother

\algrenewcommand\algorithmicend{\strut\textbf{end}}
\algrenewcommand\algorithmicdo{\strut\textbf{do}}
\algrenewcommand\algorithmicwhile{\strut\textbf{while}}
\algrenewcommand\algorithmicfor{\strut\textbf{for}}
\algrenewcommand\algorithmicforall{\strut\textbf{for all}}
\algrenewcommand\algorithmicloop{\strut\textbf{loop}}
\algrenewcommand\algorithmicrepeat{\strut\textbf{repeat}}
\algrenewcommand\algorithmicuntil{\strut\textbf{until}}
\algrenewcommand\algorithmicprocedure{\strut\textbf{procedure}}
\algrenewcommand\algorithmicfunction{\strut\textbf{function}}
\algrenewcommand\algorithmicif{\strut\textbf{if}}
\algrenewcommand\algorithmicthen{\strut\textbf{then}}
\algrenewcommand\algorithmicelse{\strut\textbf{else}}

\algrenewcommand\algorithmicrequire{\strut\textbf{Input:}}
\algrenewcommand\algorithmicensure{\strut\textbf{Output:}}

\let\oldState\State
\renewcommand{\State}{\oldState\strut}

\usepackage{array}

\usepackage{url}

\NewEnviron{resizealign}{\sbox0{
    $\begin{matrix}\displaystyle\BODY\end{matrix}$}%
  \sbox1{$(\theequation)$}%
  \sbox2{\parbox{\dimexpr \wd0 + 2\wd1}%
    {\begin{align}\BODY\end{align}}}
  \noindent\resizebox{\columnwidth}{!}{\usebox2}%
}

\begin{document}

\title{A~Min-Max~Fair~Resource~Allocation~Framework~for Optical x-haul and DU/CU in Multi-tenant O-RANs}

\author{\IEEEauthorblockN{Sourav Mondal and Marco Ruffini}
\IEEEauthorblockA{CONNECT Centre for Future Networks and Communication, Trinity College Dublin, Ireland \\
\texttt{somondal@tcd.ie, marco.ruffini@tcd.ie}}
}

\maketitle
\begin{abstract}
The recently proposed open-radio access network (O-RAN) architecture embraces cloudification and network function virtualization techniques to perform the base-band function processing by dis-aggregated radio units (RUs), distributed units (DUs), and centralized units (CUs). This enables the cloud-RAN vision in full, where mobile network operators (MNOs) could install their own RUs, but then lease on-demand computational resources for the processing of DU and CU functions from commonly available open-cloud (O-Cloud) servers via open x-haul interfaces due to variation of load over the day. This creates a multi-tenant scenario where multiple MNOs share networking as well as computational resources. In this paper, we propose a framework that dynamically allocates x-haul and DU/CU resources in a multi-tenant O-RAN ecosystem with min-max fairness guarantees. This framework ensures that a maximum number of RUs get sufficient resources while minimizing the OPEX for their MNOs. Moreover, in order to provide an access network architecture capable of sustaining low-latency and high capacity between RUs and edge-computing devices, we consider time-wavelength division multiplexed (TWDM) passive optical network (PON)-based x-haul interfaces where the PON virtualization technique is used to provide a direct optical connection between end-points. This creates a virtual mesh interconnection among all the nodes such that the RUs can be connected to the Edge-Clouds at macro-cell RU locations as well as to the O-Cloud servers at the central office locations. Furthermore, we analyze the system performance with our proposed framework and show that MNOs can operate with a better cost-efficiency than baseline greedy resource allocation with uniform cost-sharing.
\end{abstract}

\vspace{-0.5\baselineskip}
\section{Introduction} \label{sec1}
The fifth-generation (5G) radio access networks (RANs) are standardized to meet a diverse set of QoS requirements to support broadband, low-latency, and machine type communications. Applications like mixed reality, telesurgery, high-definition video streaming, Internet-of-Things, to name a few, will be free from the spectrum crunch and network resource scarcity issues of the legacy RANs. However, the existing 5G architectures lack sufficient flexibility and intelligence for efficient management of such requirements \cite{6G_vision}. Therefore, the necessity for a major architectural revolution is envisaged for beyond 5G and sixth-generation (6G) RANs. Over the past few years, major mobile network operators (MNOs) across the globe are collaborating within the \emph{Open-RAN (O-RAN) Alliance} to standardize an open and smart RAN architecture that can perform complex RAN management with the aid of software-defined networking (SDN), network function virtualization (NFV), and edge computing (EC) technologies \cite{oran}. In this architecture, 3GPP recommends RUs to perform low-PHY functions (Split 7.1/2/3), while high-PHY, MAC, RLC, RRC, and PDCP functions are processed by the DU/CUs, hosted on O-Clouds with commercial off-the-shelf (COTS) hardware. Recently, the IEEE P1914.1 standard body has been formed for specifying the next generation front-haul interface (NGFI). The RU-DU interface is known as the \emph{NGFI-I} or \emph{front-haul} (one-way latency $\leq 100$ $\mu$sec) and the DU-CU interface is known as the \emph{NGFI-II} or \emph{mid-haul} (one-way latency $\leq 1-10$ msec) \cite{ngfi2}. The interface beyond CU to the 5G core is known as the \emph{back-haul} and hence, the general term \emph{x-haul} is used.\par
The incorporation of O-Clouds for DU/CU function processing over open x-haul interfaces in the O-RAN architecture creates new business opportunities for small, medium, and large MNOs as well as cloud infrastructure providers \cite{aceg1}. In turn, a \emph{multi-tenant O-RAN ecosystem} is created where several MNOs deploy their RUs with macro and small-cell coverage over a certain geographic area, but procure x-haul and DU/CU function processing resources from the open and shared resource pool \cite{smsng}. The primary benefit of this multi-tenant O-RAN architecture is minimization of the CAPEX and OPEX for the MNOs. The techno-economic analysis in \cite{tech_eco} shows that $\sim40\%$ CAPEX and $\sim15\%$ OPEX over 5 years can be reduced. In practice, a neutral mediator like the government, municipality, or an alliance of MNOs owns the open x-haul and O-Cloud resources and decides the resource distribution among the MNOs according to their demands. Furthermore, a competitive market model can also be created where the MNOs compete against each other or form coalitions for procuring their required x-haul and O-Cloud resources.\par
The authors of \cite{tenant1} demonstrated a multi-vendor multi-standard PON for 5G x-haul that performs the control and management operations by SDN/NFV technologies. In \cite{tenant2}, the authors demonstrated a virtual network controller enabled multi-tenant virtual network on top of multi-technology optical transport networks. Moreover, the authors of \cite{tenant3} proposed a network slicing-enabled dynamic resource allocation framework for multi-tenant 5G transport networks. The authors of \cite{tenant4} proposed an on-demand cooperative network infrastructure sharing framework by adopting complex network theory in a multi-tenant environment. The authors of \cite{NS_game} proposed a game-theoretic resource allocation mechanism for network slicing-enabled multi-tenant mobile networks. However, to the best of our knowledge, the resource allocation problem for multi-tenant O-RAN ecosystems is yet to be investigated.\par
The O-Clouds are generally installed at central office (CO) locations, but their significant intermediate distance becomes disadvantageous for supporting low-latency applications (especially for the ultra-low latency front-haul interface). This hurdle can be overcome by installing Edge-Clouds at macro-cell RU locations and hosting the DU/CU and cores functions locally for some of the neighboring small-cell RUs. The RUs supporting latency-tolerant broadband applications can be connected to O-Cloud and 5G core without such issues. Therefore, we consider the TWDM-PON architecture proposed in \cite{sandip} as the x-haul interfaces because it can create a logical mesh topology to facilitate the small-cell RUs to be connected with O-Clouds at CO locations or Edge-Clouds at macro-cell RU locations in a flexible manner. This architecture \emph{supports East-West communication through PON virtualization} along with traditional North-South communication and its efficiency over similar architectures in literature was also shown.\par
We observe that connecting RUs from different MNOs to either Edge-Cloud or O-Cloud over open x-haul interfaces can be classified as a \emph{resource allocation problem}, which frequently arises in various network scenarios. Some popular methods to address such problems are uniform sharing, utility maximization, max-min/min-max fairness, and proportional fairness \cite{fair}. In the uniform sharing approach, the total cost of all the resources are uniformly distributed among the RUs, but this may be inefficient due to non-uniform resource demands of the RUs. In the utility maximization approach, resources are allocated among MNOs to maximize their economic profit or throughput, but this approach might allocate a high amount of resources to the wealthier MNOs while causing starvation to others. On the other hand, in the min-max fairness approach (dual of max-min), resources are allocated to RUs such that the maximum OPEX of the RUs are minimized while satisfying their QoS requirements. The proportional fairness approach is very similar to the max-min fairness approach, but maximizes a logarithmic utility function for resource allocation. Therefore, in this work, \emph{we embrace the min-max fairness approach to allocate open x-haul and Edge/O-Cloud resources in a multi-tenant O-RAN ecosystem}. Our primary contributions in this paper can be summarized as follows:
\vspace{-0.2\baselineskip}
\begin{enumerate}[(a)]
\item We propose a multi-tenant O-RAN architecture where the RUs from multiple MNOs can be flexibly connected to Edge/O-Clouds for their DU/CU and core functions over TWDM-PON-based x-haul interfaces via East-West and North-South communication links.
\item We formulate an integer non-linear programming problem (INLP) for the min-max fair resource allocation. In this formulation, we minimize the cost for leasing networking and computational resources of each RU while ensuring the QoS requirements of the low-latency and broadband applications are satisfied.
\item We also design a time-efficient algorithm for practical implementation of the framework. Through numerical evaluation, we analyze the characteristics of our proposed framework and show that it is cost-efficient than resource allocation with uniform cost-sharing.
\end{enumerate}
\par The rest of this paper is organized as follows. Section \ref{sec2} describes the multi-tenant O-RAN architecture. Section \ref{sec3} presents the system model. Section \ref{sec4} formulates the min-max fair resource allocation problem and designs an algorithm. Section \ref{sec5} presents numerical evaluations. Finally, Section \ref{sec6} summarizes the achievements of the proposed framework.

\begin{figure*}[t!]
\centering
\includegraphics[width=0.9\textwidth,height=10.35cm]{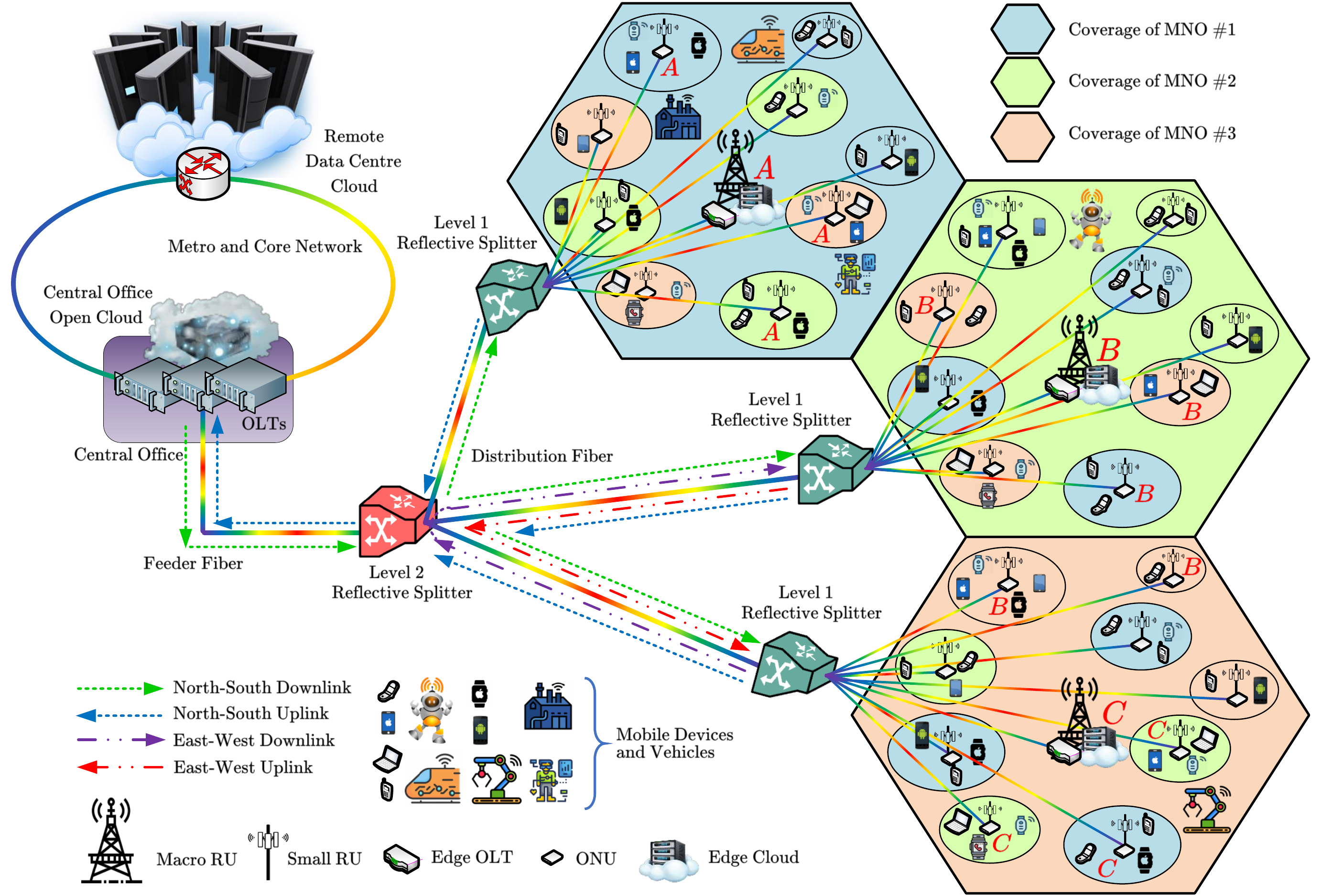}
\caption{The proposed TWDM-PON-based multi-tenant O-RAN architecture where RUs from multiple MNOs (hexagonal macro-cell and circular small-cell coverage area are shown in blue, green, and orange for three different MNOs) can be connected to Edge-Cloud or CO O-Cloud via the North-South or East-West virtual-PONs (indicated by red A, B, C) for the respective DUs and CUs.}
\label{architecture}
\end{figure*}
\setlength{\textfloatsep}{1pt}

\section{Multi-tenant O-RAN Architecture} \label{sec2}
Fig. \ref{architecture} shows the considered TWDM-PON-based x-haul interfaces for 5G O-RAN architecture in a multi-tenant scenario. Recently, ITU-T has drafted recommendations for use of TWDM-PONs as an optical x-haul solution supporting small cell connectivity. Moreover, other recent work has addressed PON slicing isolation \cite{marco1} and compliance with service level agreement (SLAs) \cite{marco2}. Furthermore, TWDM-PONs could support 100 Gbps or more aggregated datarate (i.e., in upcoming standardization) which can be scaled further by combining additional wavelengths. All the RUs are connected to a CO via a TWDM-PON with multi-level \emph{reflective splitters}. These are designed so that they can be dynamically reconfigured to pass through or reflect back (i.e., towards the end points) desired set of wavelengths (the concept is taken from \cite{sandip}). A set of level-1 reflective splitters are used to connect RUs and edge cloud nodes directly, while multiple level-1 reflective splitters are connected to a level-2 reflective splitter to reach through other PON branches. The traditional North-South communications links between the optical line terminals (OLTs) at CO and ONUs are shown in green (downlink) and blue (uplink).\par 
The local connectivity between small cell RUs and macro-cell RU is achieved by installing an Edge-OLT at the macro-cell RU location, while the small cell can host simpler ONU type of device. Control signals via the North-South communication links can be sent to these Edge-OLTs and ONUs to create virtual-PON instances that communicate via the East-West communications links as shown in red (downlink) and orange (uplink). For example, three virtual-PON instances are shown in Fig. \ref{architecture} and the ONUs and Edge-OLTs belonging to the same virtual-PON are labelled as A, B, and C in red. Note that ONUs in virtual-PON instance A communicate only via level-1 reflective splitter. The same occurs for instance C. However, ONUs in virtual-PON instance B can communicate via both level-1 and level-2 reflective splitters (i.e., they extend across two PON branches). The O-Cloud at CO location hosts the DU/CU and 5G core functions for the connected RUs. However, the Edge-Clouds can host the DU/CU and local core functions, especially to support low-latency applications. The back-haul traffic can be routed to the remote data centres over metro and core networks.\par
In a multi-tenant O-RAN ecosystem, multiple MNOs install neighboring RUs with hexagonal macro-cell and circular small-cell coverage, as shown in Fig. \ref{architecture}. We consider that each MNO pays a fee for leasing of networking (i.e., for x-haul) and computing resources (i.e., DU/CU processing at Edge/O-Clouds). We also assume that leasing fees are reduced where the Edge-Cloud node is owned by the same MNO. However, all the MNOs need to pay a default price to the mediator as the open platform provider and cost of the resources required for RAN management and control plane operations.

\section{System Model} \label{sec3}
In this section, we describe the TWDM-PON-based x-haul communication and RU-DU-CU function processing models considered for our problem formulation.\par
The required datarate for the x-haul interface mainly depends on the split option chosen between RU and DU. With Split-7.2, all the radio frequency processing, fast Fourier transform (FFT)/inverse FFT, cyclic prefix removal/addition, digital beamforming, and resource element mapping are done at the RU. The datarate can be calculated as follows:
\begin{align}
    W_{7.2} = N_P \times N_{RB} \times N_{RB}^{SC} \times N^{SF}_{sym} \times T^{-1}_{SF} \times \mu \times N_Q \times 2 \times \zeta, \label{eq01}
\end{align}
where, $N_P$ denotes the number of antenna ports, $N_{RB}$ denotes the number of resource blocks (RB), $N_{RB}^{SC}$ denotes the number of sub-carriers per RB, $T^{-1}_{SF}$ denotes sub-frame duration, $\mu$ denotes the maximum RB utilization, $N_Q$ denotes the quantizer bit resolution per I/Q dimension, and $\zeta$ denotes the front-haul overhead. With Split-7.3, precoding, layer mapping, and modulation are also done with the aforementioned tasks that reduces the datarate further and calculated as follows:
\begin{align}
    W_{7.3} = N_{Layer} \times &N_{RB} \times N^{RB}_{SC} \times N^{SF}_{sym} \times T^{-1}_{SF} \nonumber\\
    &\times \mu \times (1-\eta) \times N_Q \times \log_2(M_{mod}) \times \zeta, \label{eq02}
\end{align}
where, $N_{Layer}$ denotes the number of spatial layers, $\eta$ denotes resource overhead, and $M_{mod}$ denotes the modulation order.\par
The x-haul data are transmitted as multiple periodic bursts of Ethernet frames within each transmission time interval (TTI) over TWDM-PON and \emph{cooperative dynamic bandwidth allocation (Co-DBA)} protocol is used. The burst interval is denoted by $\delta_t$ (sec) and is equal to the average sub-frame duration or TTI. The number of frames in a burst, denoted by $\mathcal{B}$, can be calculated as $\mathcal{B} = \lceil R_D\times \delta_t/\mathcal{P}\rceil$, where $R_D$ denotes required data rate and $\mathcal{P}$ denotes the payload size of an Ethernet frame (1500 Bytes). Hence, the actual throughput of a flow can be calculated by $(\mathcal{B}\times F/\delta_t)$, where $F$ is the maximum Ethernet frame size (1542 Bytes).\par
The total RU/DU/CU functions processing effort per TTI in giga operations per second (GOPS) is given by \cite{mgain}:
\begin{align}
    \mathcal{C}_{RDC} = \left(3N_{ant} + N_{ant}^2 + \frac{1}{3} \times \mathcal{M}\times \Psi\times N_{Layer}\right)\times \frac{N_{RB}}{5}, \label{eq03}
\end{align}
where, $N_{ant}$ denotes the number of MIMO antennas, $\mathcal{M}$ denotes the number of modulation bits, and $\Psi$ denotes the coding rate. This total computational effort $\mathcal{C}_{RDC}$ is distributed among RU, DU, and CU based on the chosen intermediate split options. For example, 40\% processing is done by RU with Split-7.2 but 50\% processing is done by RU with Split-7.3. The remainder of the processing is done by the DU/CU and the total RU/DU/CU processing time can be computed by the polynomial expression provided in \cite{F_split}.

\section{Min-Max Fair Resource Allocation Problem} \label{sec4}
In this section, we formulate an integer non-linear programming (INLP) for connecting RUs from different MNOs either to an Edge-Cloud or an O-Cloud over TWDM-PON-based x-haul interfaces where their respective DUs and CUs can be hosted. we denote the set of RUs by $\mathcal{R} = \{1,2,\dots,R\}$, the set of Edge-Clouds by $\mathcal{E} = \{1,2,\dots,E\}$, and the set of O-Clouds by $\mathcal{Q} = \{1,2,\dots,Q\}$. The binary variable $x_{re}$ denotes if an RU $r\in\mathcal{R}$ is connected to an Edge-Cloud $e\in\mathcal{E}$ and $x_{rq}$ denotes if an RU $r\in\mathcal{R}$ is connected to an O-Cloud $q\in\mathcal{Q}$. The parameters $z_{re}$ and $z_{rq}$ denote if RU $r$ and Edge-Cloud $e$ and RU $r$ and O-Cloud $q$ can be connected over a virtual-PON when their values are 1. The parameters $C_r$, $C_{\lambda}$, and $C_P$ denote the default cost to the mediator, cost for throughput used, and cost for GOPS leased for each RU $r$, respectively. The cost-discount parameter $\gamma_{re} = 0.5$ if RU $r$ and Edge-Cloud $e$ belong to the same MNO, otherwise $\gamma_{re} = 1$. The parameters $W_r^{UL}$ and $W_r^{DL}$ denote the required uplink and downlink datarate of RU $r$. The parameters $B_e^{UL}$, $B_e^{DL}$ denote the maximum uplink and downlink datarate of East-West TWDM-PON links and $B_q^{UL}$ and $B_q^{DL}$ denote the maximum uplink and downlink datarate of North-South TWDM-PON links, but can vary according to the number of configured wavelengths. The parameters $\eta_r^{UL}$ and $\eta_r^{DL}$ denote the required uplink and downlink GOPS/TTI, $H_r^{UL}$ and $H_r^{DL}$ denote the available uplink and downlink GOPS/TTI for RU processing. The parameters $\Gamma_r^{UL}$ and $\Gamma_r^{DL}$ denote the required GOPS/TTI for DU/CU processing of RU $r$ and $G_e^{UL}$, $G_e^{DL}$, $G_q^{UL}$, $G_q^{DL}$ denote maximum available GOPS/TTI at Edge-Clouds $e$ and O-Clouds $o$. The parameters $\theta_{re}$ and $\theta_{rq}$ denote the burst interval over which data are transmitted from ONUs connected to RU $r$ in East-West or North-South TWDM-PONs. Finally, the parameters $\Delta_r^{H}$ and $\Delta_r^{RDC}$ denote the maximum tolerable x-haul latencies and total RU/DU/CU processing for RU $r$. Now, we formulate the min-max fair resource allocation problem as follows:
\begin{align}
    &\mathcal{P}: \min_{x_{re},x_{rq}}\max_r \left\{C_r + C_\lambda \left(\sum_{e\in\mathcal{E}} B_{re} x_{re} +\sum_{q\in\mathcal{Q}} B_{rq} x_{rq} \right) \right.\nonumber\\
    & \left.\quad\quad\quad\quad\quad\quad\quad + C_P \left(\sum_{e\in\mathcal{E}} \gamma_{re} G_{re} x_{re} +\sum_{q\in\mathcal{Q}} G_{rq} x_{rq} \right) \right\}, \label{eq04}\\
    & \text{subject to} \quad x_{re} \leq z_{re}, \forall r\in\mathcal{R}, e\in\mathcal{E}, \label{eq05}\\
    & \quad\quad\quad\quad\quad x_{rq} \leq z_{rq}, \forall r\in\mathcal{R}, q\in\mathcal{Q}, \label{eq06}\\
    & \quad\quad\quad\quad\quad \sum_{e\in\mathcal{E}} x_{re} + \sum_{q\in\mathcal{Q}}x_{rq} \leq 1, \forall r\in\mathcal{R}, \label{eq07}
\end{align}
\vspace{-1.2\baselineskip}
\begin{align}
    B_{re} &= \left(\frac{W_r^{UL} B_e^{UL}}{\epsilon+\sum_r x_{re}W_r^{UL}}\right)  + \left(\frac{W_r^{DL} B_e^{DL}}{\epsilon+\sum_r x_{re}W_r^{DL}}\right), \forall r, e, \label{eq08}\\
    B_{rq} &= \left(\frac{W_r^{UL} B_q^{UL}}{\epsilon+\sum_r x_{rq}W_r^{UL}}\right)  + \left(\frac{W_r^{DL} B_q^{DL}}{\epsilon+\sum_r x_{rq}W_r^{DL}}\right), \forall r, q, \label{eq09}\\
    G_{re} &= \left(\frac{\Gamma_r^{UL} G_e^{UL}}{\epsilon+\sum_r x_{re} \Gamma_r^{UL}}\right)  + \left(\frac{\Gamma_r^{DL} G_e^{DL}}{\epsilon+\sum_r x_{re} \Gamma_r^{DL}}\right), \forall r, e, \label{eq10}\\
    G_{rq} &= \left(\frac{\Gamma_r^{UL} G_q^{UL}}{\epsilon+\sum_r x_{rq} \Gamma_r^{UL}}\right)  + \left(\frac{\Gamma_r^{DL} G_q^{DL}}{\epsilon+\sum_r x_{rq} \Gamma_r^{DL}}\right), \forall r, q. \label{eq11}
\end{align}
\par The objective function of the problem $\mathcal{P}$ is given by (\ref{eq04}). The constraint (\ref{eq05})-(\ref{eq06}) ensures that RU $r$ can be associated with Edge-Cloud $e$ or O-Cloud $o$ only when a virtual-PON connection exists and the constraint (\ref{eq07}) restricts RU $r$ to be connected to either Edge-Cloud $e$ or O-Cloud $o$ (only one at max). The constraints (\ref{eq08})-(\ref{eq09}) indicate the allocated throughput to RU $r$ over x-haul interface to Edge-Cloud $e$ or O-Cloud $o$ ($\epsilon \approx 0$). Similarly, the constraints (\ref{eq10})-(\ref{eq11}) indicate the allocated GOPS to RU $r$ for DU/CU processing at Edge-Cloud $e$ or O-Cloud $o$. In addition, the following latency constraints are also required to be considered $\forall r\in\mathcal{R}, e\in\mathcal{E}, q\in\mathcal{Q}$:
\begin{align}
    & x_{re}\left\{\delta_{re} +\frac{D_{re}}{v_c} \right\} + {\left\lceil \frac{\theta_{TTI}}{\theta_{re}} \right\rceil} \left\{\frac{\sum_r x_{re} W_r^{UL}\theta_{re}}{B_e^{UL}} \right\} \nonumber\\
    & + x_{rq}\left\{\delta_{rq} +\frac{D_{rq}}{v_c} \right\} + {\left\lceil \frac{\theta_{TTI}}{\theta_{rq}} \right\rceil} \left\{\frac{\sum_r x_{rq} W_r^{UL}\theta_{rq}}{B_q^{UL}} \right\} \leq \Delta_r^{H}, \label{eq12} \\
    & x_{re}\left\{\frac{D_{re}}{v_c} \right\} + {\left\lceil \frac{\theta_{TTI}}{\theta_{re}} \right\rceil} \left\{\frac{\sum_r x_{re} W_r^{DL}\theta_{re}}{B_e^{DL}} \right\} \nonumber\\
    & + x_{rq}\left\{\frac{D_{rq}}{v_c} \right\} + {\left\lceil \frac{\theta_{TTI}}{\theta_{rq}} \right\rceil} \left\{\frac{\sum_r x_{rq} W_r^{DL}\theta_{rq}}{B_q^{DL}} \right\} \leq \Delta_r^{H}, \label{eq13} \\
    &\quad \frac{\eta_r^{UL}}{H_r^{UL}} + \left\{\frac{\sum_r x_{re} \Gamma_r^{UL}}{G_e^{UL}} \right\} + \left\{\frac{\sum_r x_{rq} \Gamma_r^{UL}}{G_q^{UL}} \right\} \leq \frac{\Delta_r^{RDC}}{\theta_{TTI}}, \label{eq14} \\
    &\quad \frac{\eta_r^{DL}}{H_r^{DL}} + \left\{\frac{\sum_r x_{re} \Gamma_r^{DL}}{G_e^{DL}} \right\} + \left\{\frac{\sum_r x_{rq} \Gamma_r^{DL}}{G_q^{DL}} \right\} \leq \frac{\Delta_r^{RDC}}{\theta_{TTI}}. \label{eq15}
\end{align}
\par The constraint (\ref{eq12}) ensures that the uplink x-haul latency from RU $r$ to Edge-Cloud $e$ or O-Cloud $q$ is within $\Delta_r^{H}$. The parameters $\delta_{re}$ and $\delta_{rq}$ denote the \emph{reduced average queuing latency} of uplink data due to Co-DBA at ONUs. The second terms with $x_{re}$ and $x_{rq}$ indicate \emph{propagation latency} where the parameters $D_{re}$ and $D_{rq}$ denote the distance from RU $r$ to Edge-Cloud $e$ and O-Cloud $q$ and $v_c$ denotes speed of light within fiber ($2\times10^5$ km/s). The third terms with $x_{re}$ and $x_{rq}$ indicate \emph{data transmission latency} where data are transmitted in multiple bursts of duration $\theta_{re}$ and $\theta_{rq}$ within each TTI $\theta_{TTI}$. Similarly, the constraint (\ref{eq13}) ensures that the downlink x-haul latency from RU $r$ to Edge-Cloud $e$ or O-Cloud $q$ is within $\Delta_r^{H}$. Finally, constraints (\ref{eq14})-(\ref{eq15}) indicate the RU/DU/CU processing latencies are within $\Delta_r^{RDC}$. The first term indicates the RU processing latency and the second and third terms indicate the DU/CU processing latencies at Edge-Cloud $e$ and O-Cloud $o$, respectively.\par
Now, we design a solution algorithm based on the approach in \cite{mx_mn_GB}. By careful observation, we can understand that cost of each RU in objective (\ref{eq04}) can be minimized if each x-haul link and Edge/O-Cloud resources are leased by a maximum number of RUs while satisfying constraints (\ref{eq08})-(\ref{eq15}). Let us denote $\tilde{\mathcal{R}}$ as the set of all unassigned RUs and \emph{the RU indices are mixed according to the ownership percentage of the MNOs to ensure fairness}. Initially, a maximum number of RUs are connected to their nearest Edge/O-Cloud, i.e., $y = \arg \min_y \{D_{ry}\}, y\in(\mathcal{E}\cup\mathcal{Q})$ subject to constraints (\ref{eq05})-(\ref{eq07}), (\ref{eq12})-(\ref{eq15}) are satisfied while prioritizing Edge-Clouds from the same MNO, i.e., $\gamma_{re} = 0.5$. In the following iterations, we check if the RU $r$ with maximum OPEX can be reallocated to a different Edge-Cloud $e$ or O-Cloud $q$ such that its OPEX can be reduced to any other RU with similar resource requirements while the constraints (\ref{eq12})-(\ref{eq15}) are not violated. If impossible, then we remove $r$ from $\tilde{\mathcal{R}}$, i.e., $\tilde{\mathcal{R}} \gets \tilde{\mathcal{R}}\setminus \{r\}$. We continue this process until $\tilde{\mathcal{R}} = \emptyset$ and the loop terminates. If all the RUs could not be connected to Edge/O-Clouds due to saturation of x-haul and/or Edge-Cloud resources, then the unassigned RUs will suffer from resource outage. Now, we analyze that the complexity of the loop (1-4) is $\mathcal{O}(|\mathcal{R}|\times (|\mathcal{E}|+|\mathcal{Q}|))$ and the worst-case complexity of the loop (5-14) is $\mathcal{O}(|\mathcal{R}|^2 \times (|\mathcal{E}|+|\mathcal{Q}|))$. Thus, the overall complexity of Algorithm \ref{alg1} is given by $\mathcal{O}(|\mathcal{R}|^2 \times (|\mathcal{E}|+|\mathcal{Q}|))$.

\begin{algorithm}[b!]
\caption{Algorithm for min-max fair resource allocation} \label{alg1}
\hspace*{\algorithmicindent} \textbf{Input:} $\mathcal{R}, \mathcal{E}, \mathcal{Q}, D_{re}, D_{rq}, B_e^{U/DL}, B_q^{U/DL}, W_r^{U/DL}, \Gamma_r^{U/DL}$\\
\hspace*{\algorithmicindent} \textbf{Output:} Near-optimal solution: $x_{re}^*$ and $x_{rq}^*$\\
\hspace*{\algorithmicindent} \textbf{Initialize:} $\tilde{\mathcal{R}} = \mathcal{R}$ and mix by the ownership of MNOs;
\begin{algorithmic}[1]
\For{$r \leftarrow 1$ \textbf{to} $|\mathcal{R}|$}
    \State Find $\min_y \{D_{ry}\}, y\in(\mathcal{E}\cup\mathcal{Q})$ s.t. (\ref{eq05})-(\ref{eq07}), (\ref{eq12})-(\ref{eq15});
    \State Set $x_{re}\gets 1$ or $x_{rq} \gets 1$; \Comment{\textit{priority to $\gamma_{re} = 0.5$}}
\EndFor
\While{$\tilde{\mathcal{R}} \neq \emptyset$}
    \For{$r \leftarrow 1$ \textbf{to} $|\tilde{\mathcal{R}}|$} \Comment{\textit{start from $r$ with max cost}}
        \State Find all $y\in(\mathcal{E}\cup\mathcal{Q})$ that reduces cost of $r$;
        \If{cost of $r$ can be reduced to another RU with similar resource demand and not violating (\ref{eq12})-(\ref{eq15})}
            \State Reassign $r$ to this new $e$ or $q$;
        \Else
            \State Set $\tilde{\mathcal{R}} \gets \tilde{\mathcal{R}}\setminus \{r\}$; \Comment{\textit{remove $r$ from $\tilde{\mathcal{R}}$}}
        \EndIf
    \EndFor
\EndWhile
\State \textbf{return} $x_{re}$ and $x_{rq}$;
\end{algorithmic}
\end{algorithm}
\setlength{\textfloatsep}{5pt}

\section{Results and Discussions} \label{sec5}
To evaluate our proposed framework in Section \ref{sec4}, we consider a multi-tenant O-RAN deployment area of dimension $5\times 5$ km$^2$. In this area, 8 macro-cell RUs (coverage = 1 km) and 30 small-cell RUs (coverage = 0.5 km) from three different MNOs coexist. The x-haul datarate and RU/DU/CU processing efforts of the RUs vary over time depending on the RU configurations and the chosen split option. For example, if the RUs are configured with 2x2 MIMO, 2 layers, 50 MHz bandwidth, 15 kHz sub-carrier spacing, MCS index 16, and TTI duration = 0.5 msec, the maximum uplink datarate with Split-7.2 is 2.304 Gbps and the maximum downlink datarate with Split-7.3 is 0.432 Gbps. The total RU/DU/CU processing requirement is nearly 550 GOPS/TTI. The one-way front-haul link latency bound is 100 $\mu$sec and we choose the RU/DU/CU processing latency bound 90 $\mu$sec. The reduced queuing latency of the uplink data at the ONUs is 15 $\mu$sec as Co-DBA is used for uplink transmission and the data are transmitted as periodic bursts of 31.25 $\mu$sec. We consider two diagonally opposite corner points as the CO locations where multiple OLTs of TWDM-PONs are hosted. Each wavelength can support a throughput of 25 Gbps such that we can achieve an aggregated throughput of 100 Gbps with 4 wavelengths. A group of RU/ONUs are connected to a level-1 reflective splitter (locations found through $k$-means clustering) and a few level-1 reflective splitters are connected to a level-2 reflective splitter located at the center of the area. We assume that the RU/ONUs can be connected to the O-Clouds via the North-South links and to the Edge-Clouds via creating East-West virtual-PON links. {We arbitrarily assume that each MNO pays a default cost of \euro 100/day to the mediator. In addition, the cost for throughput used is \euro 0.5/Gbps \cite{fib_rent}, and the cost for leasing DU/CU resources is \euro 1.5/GOPS \cite{F_split}}.\par
\begin{figure}[!t]
\centering
\includegraphics[width=\columnwidth,height=5.5cm]{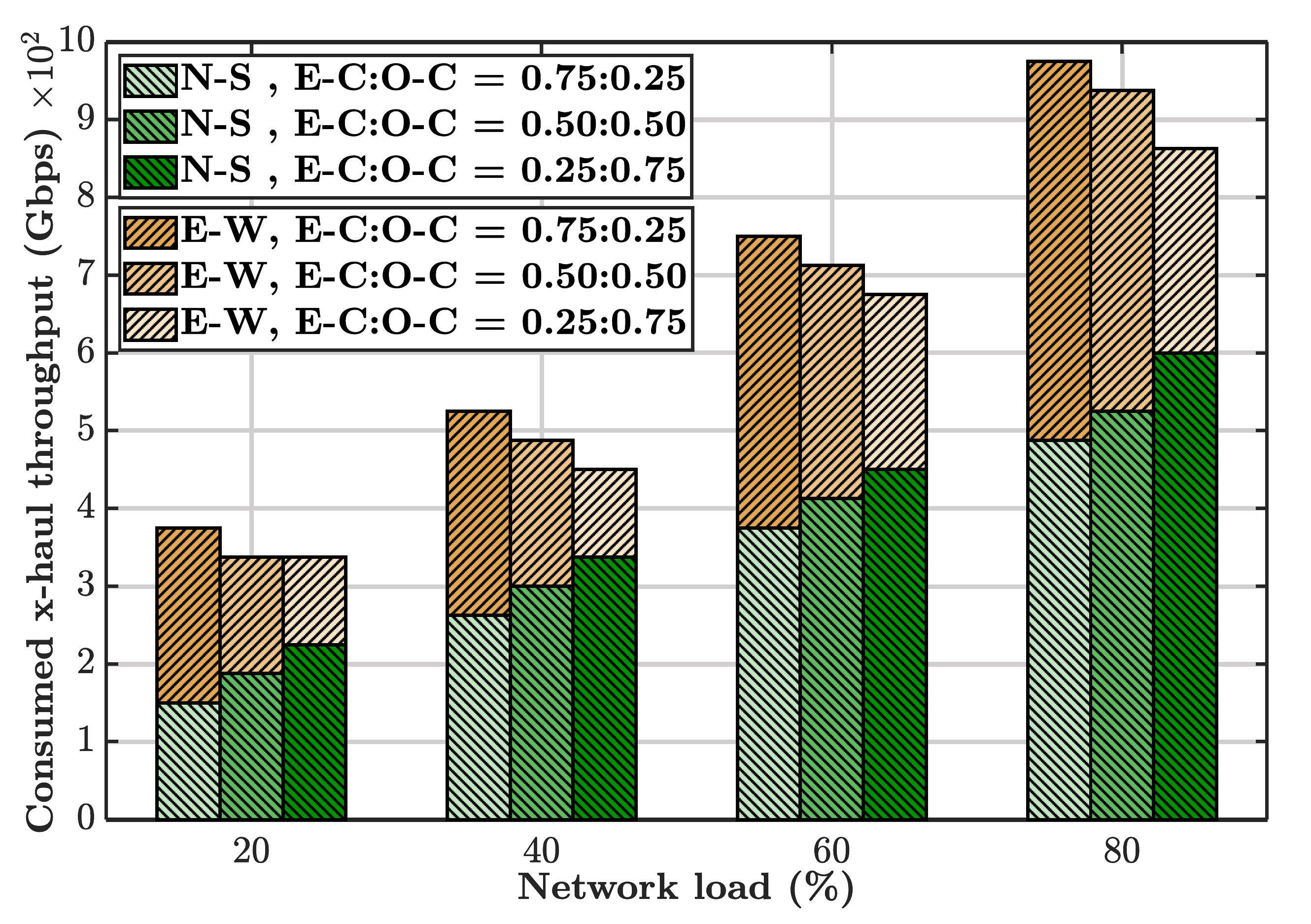}
\caption{Distribution of x-haul throughput consumption by RUs against network load over the North-South and East-West TWDM-PON links.}
\label{thpt}
\end{figure}
\setlength{\textfloatsep}{5pt}
\begin{figure}[!b]
\centering
\includegraphics[width=\columnwidth,height=5.50cm]{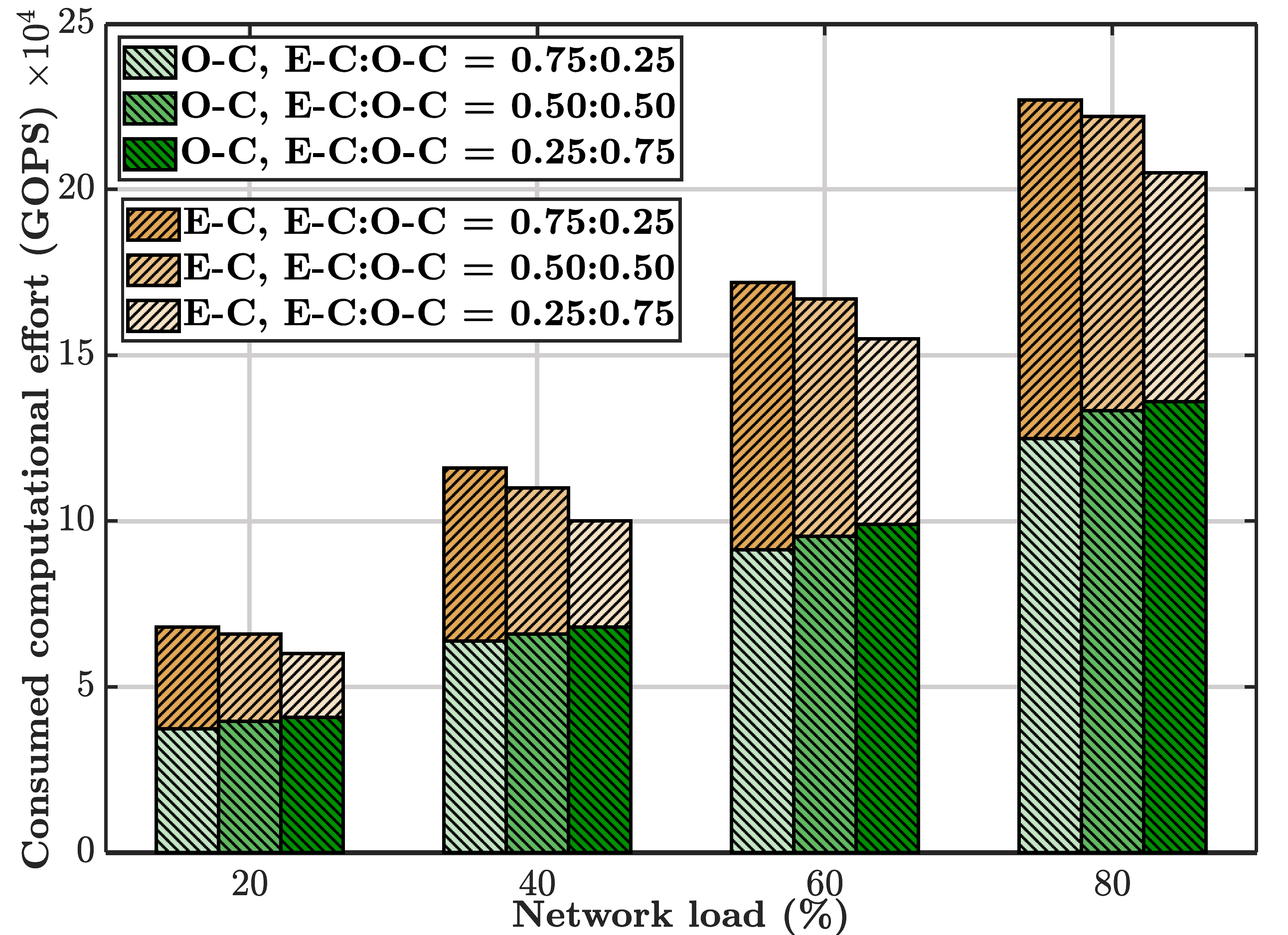}
\caption{Distribution of DU/CU processing resources leased by RUs against network load at Edge-Clouds and O-Clouds.}
\label{ruducu}
\end{figure}
\setlength{\textfloatsep}{5pt}
%
Figs. \ref{thpt}-\ref{ruducu} show the total x-haul throughput and RU/DU/CU processing resource consumption against network load (maximum load $\approx 4$ Gbps/km$^2$), respectively. {We assume that nearly 25\% low-latency applications and 75\% broadband applications are being served by the RUs}. We consider that a small MNO-1 has 25\%, a medium MNO-2 has 35\%, and a large MNO-3 has 40\% ownership of the RUs. In addition, we vary the ratio of resources in Edge-Clouds and O-Clouds as 0.25:0.75, 0.50:0.50, and 0.75:0.25 while the maximum computational effort supported available in each cloud is $10^4$ GOPS/TTI. {Both these plots indicate that the percentage of RUs connected to Edge/O-Clouds is dictated by the ratio of resources among them, i.e., when Edge-Cloud resources are higher than O-Cloud resources, a higher number of RUs are connected to the Edge-Clouds. Nonetheless, our proposed framework attempts to connect the maximum possible number of RUs to O-Clouds in all scenarios, because a higher number of RUs could be aggregated over the North-South links with a lower cost than East-West links. However, as Edge-Cloud owned by the same MNO may also minimize the cost, for some RUs (especially those serving low-latency applications) connection to the Edge-Clouds is prioritized.}\par
\begin{figure}[!t]
\centering
\includegraphics[width=\columnwidth,height=5.5cm]{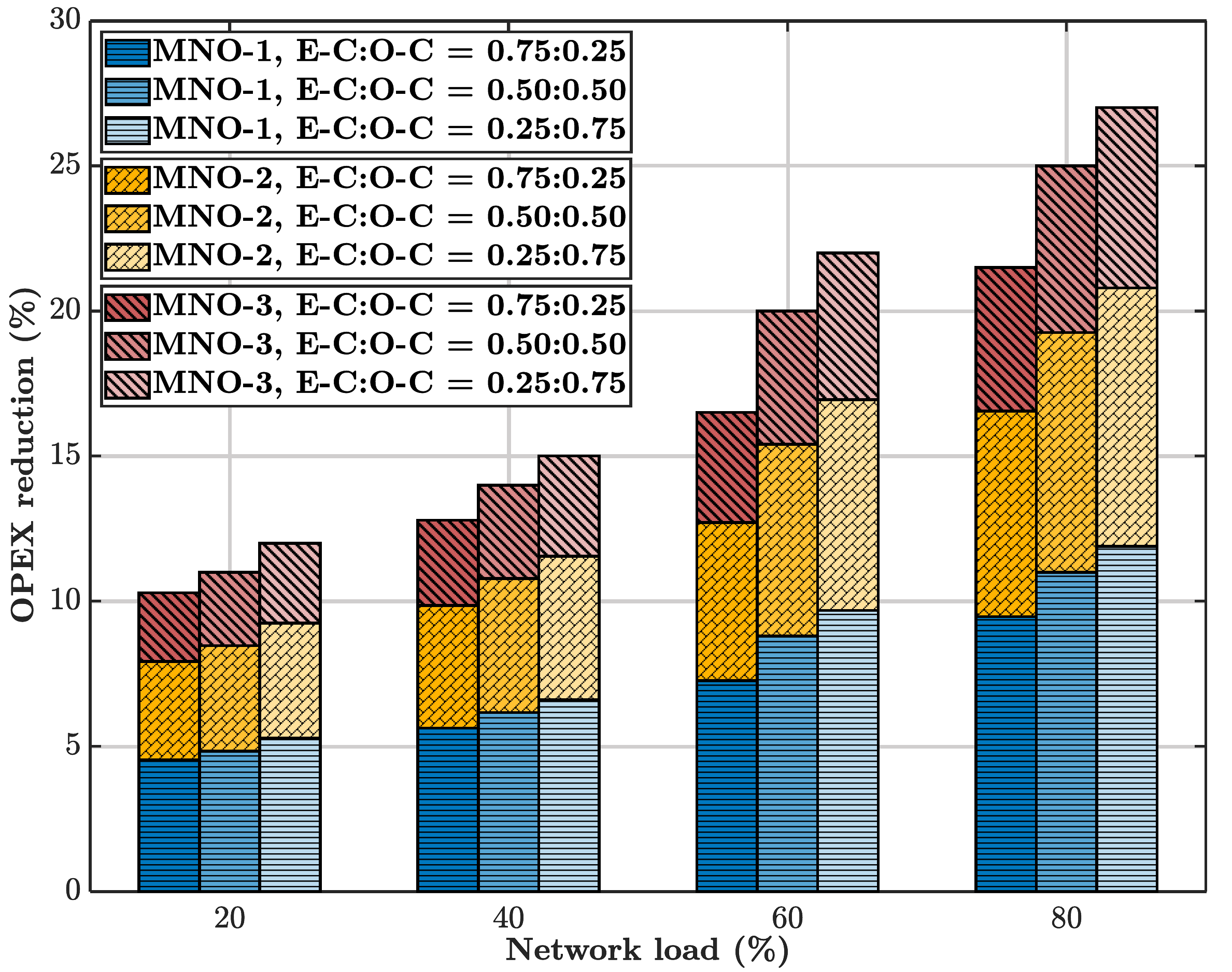}
\caption{OPEX reduction percentage of MNOs against network load by the proposed framework compared to uniform resource allocation.}
\label{cost}
\end{figure}
\setlength{\textfloatsep}{5pt}
Next, in Fig. \ref{cost} we compare the OPEX reduction percentage of multiple MNOs with our proposed framework and resource allocation with uniform cost-sharing (baseline greedy) against the network load of the considered area. With the later, sufficient x-haul and Edge/O-Cloud resources are allocated to meet the QoS requirements of all the RUs together from different MNOs and the total OPEX is uniformly shared among them. {However, this strategy often makes some of the MNOs (especially those with a smaller number of RUs) pay a higher cost for the resources than they have actually utilized, which in turn, creates unfairness in the multi-tenant O-RAN ecosystem.} In this scenario, we observe that a higher percentage of RUs from the large MNO-3 are connected to Edge-Clouds, but RUs from small MNO-1 and medium MNO-2 are mostly aggregated to O-Clouds due to a better OPEX reduction percentage. The plot shows that our proposed framework reduces the OPEX of MNOs by nearly 27\% (12\% for the small MNO-1, 9\% for the medium MNO-2, and 6\% for the large MNO-3) at 80\% network load condition. We also observe that OPEX reduction for small MNO-1 is highest when nearly 75\% of resources are placed at O-Cloud servers. These results justify that our proposed min-max fair resource allocation framework creates a multi-tenant O-RAN ecosystem that is sustainable for small, medium, as well as large MNOs.
\section{Conclusions} \label{sec6}
In this paper, we have proposed a multi-tenant O-RAN architecture where RUs from multiple MNOs can lease resources for their DU/CU functions processing from Edge/O-Clouds over TWDM-PON-based open x-haul interfaces. We have adopted a min-max fair resource allocation strategy for allocating x-haul and DU/CU processing resources to multiple MNOs. Thus, we have formulated an INLP problem and have designed a time-efficient algorithm to solve it. Through numerical evaluation, we have shown that the proposed framework reduces the OPEX of the MNOs up to 27\% than resource allocation with uniform cost-sharing (for the considered scenario, network load, and cost parameters). We strongly believe that our proposed framework is will provide important insights for exploring further O-RAN resource allocation strategies.

\section*{Acknowledgment}
Financial support from EU H2020 EDGE (grant 713567), Science Foundation Ireland (SFI) grants 17/CDA/4760 and 13/RC/2077\_P2 are gratefully acknowledged.

\bibliographystyle{IEEEtran}
\bibliography{IEEEabrv,references}

\begin{thebibliography}{10}
\providecommand{\url}[1]{#1}
\csname url@samestyle\endcsname
\providecommand{\newblock}{\relax}
\providecommand{\bibinfo}[2]{#2}
\providecommand{\BIBentrySTDinterwordspacing}{\spaceskip=0pt\relax}
\providecommand{\BIBentryALTinterwordstretchfactor}{4}
\providecommand{\BIBentryALTinterwordspacing}{\spaceskip=\fontdimen2\font plus
\BIBentryALTinterwordstretchfactor\fontdimen3\font minus
  \fontdimen4\font\relax}
\providecommand{\BIBforeignlanguage}[2]{{%
\expandafter\ifx\csname l@#1\endcsname\relax
\typeout{** WARNING: IEEEtran.bst: No hyphenation pattern has been}%
\typeout{** loaded for the language `#1'. Using the pattern for}%
\typeout{** the default language instead.}%
\else
\language=\csname l@#1\endcsname
\fi
#2}}
\providecommand{\BIBdecl}{\relax}
\BIBdecl

\bibitem{6G_vision}
W.~{Saad}, M.~{Bennis}, and M.~{Chen}, ``{A Vision of 6G Wireless Systems:
  Applications, Trends, Technologies, and Open Research Problems},'' \emph{IEEE
  Network}, vol.~34, no.~3, pp. 134--142, 2020.

\bibitem{oran}
``{O-RAN: Towards an Open and Smart RAN},'' O-RAN Alliance, Tech. Rep., Oct
  2018.

\bibitem{ngfi2}
C.-L. I \emph{et~al.}, ``{RAN Revolution With NGFI (xhaul) for 5G},'' \emph{J.
  Lightw. Technol.}, vol.~36, no.~2, pp. 541--550, 2018.

\bibitem{aceg1}
L.~Peterson \emph{et~al.}, ``{Democratizing the Network Edge},'' \emph{ACM
  SIGCOMM Comp. Commun. Rev.}, vol.~49, no.~2, p. 31–36, May 2019.

\bibitem{smsng}
\BIBentryALTinterwordspacing
``{The Open Road to 5G},'' Samsung Networks, Tech. Rep., July 2019. [Online].
  Available:
  \url{https://image-us.samsung.com/SamsungUS/samsungbusiness/pdfs/Open-RAN-The-Open-Road-to-5G.pdf}
\BIBentrySTDinterwordspacing

\bibitem{tech_eco}
B.~Naudts \emph{et~al.}, ``{Techno-economic Analysis of Software Defined
  Networking as Architecture for the Virtualization of a Mobile Network},'' in
  \emph{2012 European Workshop Softw. Defined Netw.}, 2012, pp. 67--72.

\bibitem{tenant1}
D.~Liang, R.~Gu, Q.~Guo, and Y.~Ji, ``{Demonstration of Multi-Vendor
  Multi-Standard PON Networks for Network Slicing in 5G-oriented Mobile
  Network},'' in \emph{2017 Asia Commun. Photon. Conf. (ACP)}, 2017.

\bibitem{tenant2}
R.~Vilalta \emph{et~al.}, ``{Network Virtualization Controller for Abstraction
  and Control of OpenFlow-enabled Multi-tenant Multi-technology Transport
  Networks},'' in \emph{Opt. Netw. Commun. Conf. Exhibit. (OFC)}, 2015.

\bibitem{tenant3}
M.~R. Raza \emph{et~al.}, ``{Dynamic Slicing Approach for Multi-Tenant 5G
  Transport Networks [Invited]},'' \emph{J. Opt. Commun. Netw.}, vol.~10,
  no.~1, pp. A77--A90, Jan 2018.

\bibitem{tenant4}
W.~Guan \emph{et~al.}, ``{On-Demand Cooperation Among Multiple Infrastructure
  Networks for Multi-Tenant Slicing: A Complex Network Perspective},''
  \emph{IEEE Access}, vol.~6, pp. 78\,689--78\,699, 2018.

\bibitem{NS_game}
P.~Caballero, A.~Banchs, G.~De~Veciana, and X.~Costa-Pérez, ``{Network Slicing
  Games: Enabling Customization in Multi-Tenant Mobile Networks},''
  \emph{IEEE/ACM Trans. Netw.}, vol.~27, no.~2, pp. 662--675, 2019.

\bibitem{sandip}
S.~Das \emph{et~al.}, ``{Virtualized EAST–WEST PON Architecture Supporting
  Low-latency Communication for Mobile Functional Split based on Multiaccess
  Edge Computing},'' \emph{IEEE/OSA J. Opt. Commun. Netw.}, vol.~12, no.~10,
  pp. D109--D119, 2020.

\bibitem{fair}
L.~Massoulie and J.~Roberts, ``{Bandwidth Sharing: Objectives and
  Algorithms},'' \emph{IEEE/ACM Trans. Netw.}, vol.~10, no.~3, pp. 320--328,
  2002.

\bibitem{marco1}
M.~Ruffini, A.~Ahmad, S.~Zeb, N.~Afraz, and F.~Slyne, ``{Virtual DBA:
  Virtualizing Passive Optical Networks to Enable Multi-service Operation in
  True Multi-tenant Environments},'' \emph{IEEE/OSA J. Opt. Commun. Netw.},
  vol.~12, no.~4, pp. B63--B73, 2020.

\bibitem{marco2}
F.~Slyne, S.~Zeb, and M.~Ruffini, ``{Stateful DBA Hypervisor Supporting SLAs
  with Low Latency and High Availability in Shared PON},'' in \emph{2021 Opt.
  Netw. Commun. Conf. Exhibit. (OFC)}, 2021, pp. 1--3.

\bibitem{mgain}
M.~Shehata \emph{et~al.}, ``{Multiplexing Gain and Processing Savings of 5G
  Radio-Access-Network Functional Splits},'' \emph{IEEE Trans. Green Commun.
  Netw.}, vol.~2, no.~4, pp. 982--991, 2018.

\bibitem{F_split}
Y.~Xiao \emph{et~al.}, ``{Can Fine-Grained Functional Split Benefit to the
  Converged Optical-Wireless Access Networks in 5G and Beyond?}'' \emph{IEEE
  Trans. Netw. Service Manag.}, vol.~17, no.~3, pp. 1774--1787, 2020.

\bibitem{mx_mn_GB}
D.~P. Bertsekas and R.~G. Gallager, \emph{{Flow Control in Data
  Networks}}.\hskip 1em plus 0.5em minus 0.4em\relax Prentice Hall, 1992,
  ch.~6, pp. 524--529.

\bibitem{fib_rent}
``{Preliminary Fiber Network Design and Business Plan Framework},'' Columbia
  Telecommunications Corporation, Tech. Rep., Jun 2005.

\end{thebibliography}

\end{document}